# Growth of bulk crystals of magnetic topological insulators under peritectic conditions

Anton I. Sergeev[1,2], Alexander S. Frolov[1,2], Nadezhda V. Vladimirova[1,2], Arseny A. Naumov[1], Maria Kirsanova[3], Mark A. Naumov[2], Vasily S. Stolyarov[2], Vladimir A. Golyashov[4,5], Oleg E. Tereshchenko[4,5], Ilya Klimovskikh[6], Marina E. Tamm[1], L.V. Yashina[1,2,7,*]

[1] Lomonosov Moscow State University, Leninskie Gory 1/3, Moscow 119991, Russia
[2] Center for Advanced Mesoscience and Nanotechnology, Moscow Institute of Physics and Technology, 9 Institutskiy Pereulok, National Research University, Dolgoprudny, Moscow Region 141700, Russia
[3] Center for Energy Science and Technology, Skolkovo Institute of Science and Technology, Nobel St. 3, Moscow, 121205, Russian Federation
[4] Rzhanov Institute of Semiconductor Physics, Siberian Branch, Russian Academy of Sciences, 13 Lavrentiev aven., Novosibirsk, 630090, Russia.
[5] Synchrotron Radiation Facility SKIF, Boreskov Institute of Catalysis SB RAS, 630559 Kol'tsovo, Russia
[6] Donostia International Physics Center, 20018, San Sebastian, Spain
[7] N.N. Semenov Federal Research Center for Chemical Physics, Kosygina Street 4, Moscow 119991, Russia
** Corresponding author, e-mail: yashina@inorg.chem.msu.ru*

## Abstract

Magnetic topological materials (MTIs) including $MnBi_2Te_4$ are of scientific interest due to the possibility to reveal the interplay between topological and magnetic orders. For these materials preparation issues are still challenging. In our paper we report new synthetic protocol using two-phase source in Bridgman-like growth procedure. We applied it to grow a number of TIs of $GeAs_2Te_4$ structural type including individual compounds $MnBi_2Te_4$, $MnSb_2Te_4$, $GeBi_2Te_4$, $SnBi_2Te_4$, $PbBi_2Te_4$ and several mixed crystals of whole range of compositions: $Mn(Bi,Sb)_2Te_4$, $(Ge,Mn)Bi_2Te_4$, new materials $Mn(Bi,In)_2Te_4$. The method developed can be used to obtain single crystals of any incongruently melted compounds with small region of primary crystallization from melt in the corresponding equilibrium phase diagrams.

## 1. Introduction

Nowadays, topological materials possessing robust surface or interface electronic states with certain spin texture are in focus of researches of condensed matter physics [1]. Topological insulators represent a new class of materials providing a platform for exploring exotic transport phenomena, such as the quantum anomalous Hall effect (QAHE) and the quantum spin Hall effect (QSHE). Recently, magnetic topological insulators (MTIs) have attracted considerable interests due to the possibility to study the interplay between topological and magnetic orders. Among them, the first discovered intrinsic antiferromagnetic (AFM) $Z_2$ topological insulator $MnBi_2Te_4$ has been extensively studied [2–4]. It is a layered compound composed of septuple layers (SLs) with an idealized layer sequence of Te-Bi-Te-Mn-Te-Bi-Te, where partial intermixing of the Mn and Bi positions is observed in the cation sublattice, yielding $Mn_{Bi}$ and $Bi_{Mn}$ antisite defects. $MnBi_2Te_4$ belongs to a family of topological insulators with general formula $Me^{I}Me^{II}_2Te_4$ ($Me^{I}$=Mn, Ge, Sn, Pb; $Me^{II}$=Sb, Bi) of $GeAs_2Te_4$-type structure (so called 124 compound) [5–8]. One particularly intriguing example is the case of systems with a surface-normal magnetic easy axis coexisting with a topologically nontrivial surface state. Such materials demonstrate QAHE, which manifests itself as chiral edge conduction channels that can be manipulated by switching the polarization of magnetic domains [9]. The most of studies on $MnBi_2Te_4$ deals with two aspects: realizing a high-temperature QAHE and exploring the exotic phases arising from its topological and magnetic orders, such as the axion insulator phase [10–12]. Many advances have been made in the study of $MnBi_2Te_4$, but there are still many challenges and opportunities awaiting future research [2]. First problem is to prepare high-quality samples. To tune precisely crystal composition and atomic disorders, to achieve high structural perfection, more advanced crystal growth techniques are needed, and an in-depth study on the behavior of magnetic atoms is necessary. Second, a promising possibility to tune magnetism, band structure and topology in $MnBi_2Te_4$ is alloying with nonmagnetic TIs giving rise to mixed crystals (solid solutions) with a general formula (Ge, Sn, Pb)$_x$Mn$_{1-x}$Bi$_2$Te$_4$ [5,13,14]. For such complex systems preparation methods are not yet developed. Third, there still lacks an understanding of the ARPES observations related to the magnetically opened surface gap. For transport studies higher-precision device fabrication techniques are essential in order to perform layer-resolved measurements [2].

As for sample preparation, bulk $MnBi_2Te_4$ crystals are usually grown from the melt of the stoichiometric mixture [15,16], *via* the flux method [17], or the chemical vapor transport method [18,19]. The $MnBi_2Te_4$ crystals are always *n*-doped with a carrier concentration of ~$10^{20}$cm$^{-3}$ [17,20–23]. For solid solution $Mn(Sb_xBi_{1-x})_2Te_4$, the *n*–*p* carrier transition happens at certain *x* [24]. Nevertheless, most of these demanding materials are available only in the form of thin films or tiny crystals. The synthesis of these materials in single-crystalline form from melt is a challenge due to peritectic melting and the very narrow region of primary crystallization (range of conditions providing *L-S* equilibrium) for certain compounds.

In our paper we report new synthetic protocol using two-phase source in Bridgman-like growth procedure. We applied it to grow a number of TIs of $GeAs_2Te_4$ structural type and homological structures (147, 1610-compounds) including several mixed crystals of whole range of compositions. The method developed can be useful to obtain single crystals of incongruently melted compounds with narrow region of primary crystallization from melt.

## 2. Methods

Crystals composition was characterized by X-ray fluorescence spectroscopy (XFS) using a Bruker Mistral-M1 microfocused system equipped with an XFlash 30 $mm^2$ detector; the concentrations were determined using the XSpect software by determining the area of individual peaks using the external standard model.

TEM samples were prepared using a focused ion beam technique with a Thermo Scientific Helios PFIB G4 dual beam system. Electron diffraction (ED) patterns, HAADF-STEM images and energy-dispersive X-ray (EDX) compositional maps were acquired on anaberration-corrected FEI Titan Themis Z 80-300 electron microscope operated at 200 kV and equipped with a Super-X detection system. 4–7 measurement series were taken for each crystal. The HAADF-STEM simulations were performed using the QSTEM software [25].

For XRD measurements the samples were powdered. XRD was performed using a Rigaku Smartlab SE diffractometer equipped with a PSD Pixel3D detector in the 2θ range of 10–80° with a step of 0.02° at room temperature. The data obtained were treated within Jana 2006 software package [26].

ARPES data were recorded using a SPECS GmbH ProvenX-ARPES system located in ISP SB RAS equipped with ASTRAIOS 190 electron energy analyzer with 2D-CMOS electron detector and a non-monochromated He Iα light source with $hv$=21.22 eV.

For STM measurements, crystals were cleaved under UHV conditions, at $P < 2 \times 10^{-10}$ mbar, to get clean surfaces free of atomic contamination. The samples were measured at $T$ = 4.3 K in SPECS-JT microscope. Topography images were collected in constant-current mode using electrochemically-etched tungsten tip.

Magnetic force microscopy measurements were performed using AttoDry 1000 MFM system at the temperature $T$=4.1K using Bruker MESP-LN-V2 probes covered by CoCr coating.

## 3. Results and discussion

### 3.1. Description of the three-phase method

To grow crystals of incongruently melted compound $A_mB_n$ we propose to use two-phase ($L$+$S_A$) source, where the component A has higher melting point. Our approach is illustrated in Fig. 1, where typical $T$-$x$ phase diagram is schematically presented (left) together with the corresponding temperature regime of growth (right). The compound $A_mB_n$ is formed at temperature $T_p$ according to the peritectic reaction $L$+A= $A_mB_n$. An attempt to grow crystals using melt of the exact $m:n$ stoichiometry would give rise to co-crystallization of A and $A_mB_n$, whereas the melt composition $x(L_p)$ would result in right crystal stoichiometry but also in fast melt depletion in component A and corresponding decrease of crystallization temperature according to the liquidus line. We propose to use two-phase source ($L$+$S_A$) to maintain melt composition constant and equal to $x(L_p)$ at least for any binary (or quasi-binary) systems. In this case, solid phase A gradually dissolves in melt, and thuswise the crystallization temperature is maintained constant. This approach can be easily adapted to the vertical Bridgman growth procedure. The main limitation for the system is that the component A should have lower density than the melt. In this case the solid phase A can added as compact tablet. Due to its lower density the tablet floats over the melt.

The first step of growth includes equilibration of the tablet with the melt at temperature slightly higher than the peritectic temperature $T_p$. Crystal growth proceeds at step 2 when an ampoule is pulled into negative temperature gradient. In the ideal case the pulling rate should correspond to (or less than) both the natural growth rate and the tablet dissolution rate so the crystallization front remains always at temperature $T_p$. This is illustrated in Fig. 1 (right).

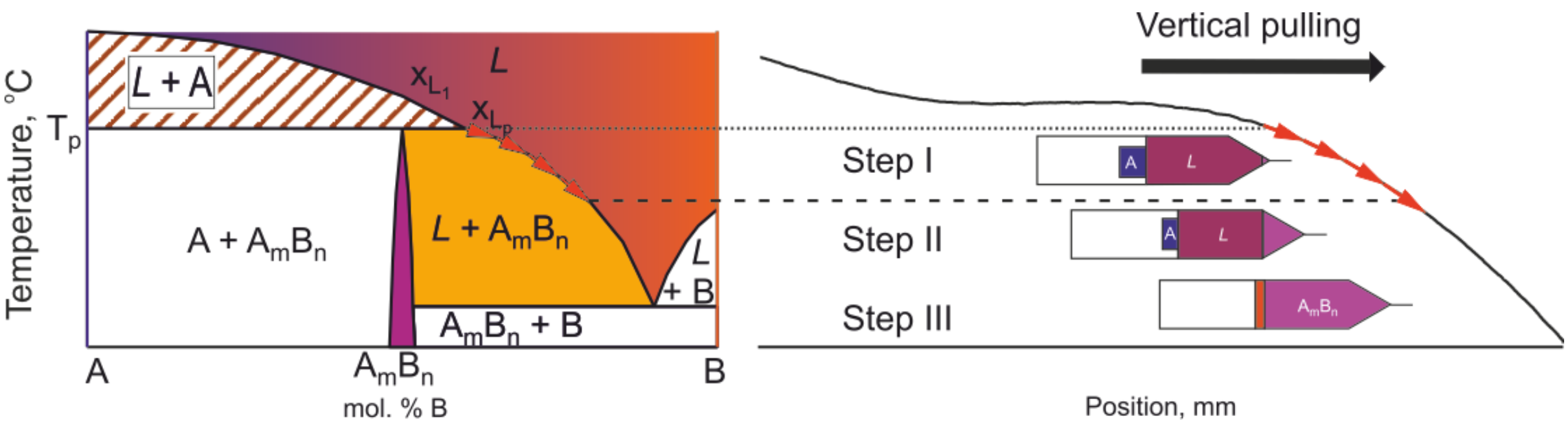

Fig.1. Schematic representation of incongruently melted compound growth using two-phase source: a fragment of *T-x* phase diagram (left) and corresponding positions of growth ampoule relative to a temperature profile during different steps (nucleation, growth, termination).

In practice it may happen that the pulling rate is higher than the natural growth rate, and the crystallization front will move into the colder zone as shown in Fig. 1 with red arrows. This region still corresponds to $A_mB_n$ crystallization until the eutectic temperature is reached. Although equilibrium conditions are violated, a local equilibrium is established at the crystallization front. In general, the natural growth rate is strongly dependent on the number of components in the system (it is lower for more components).

In our practice the nucleation temperature was determined preliminary using differential thermal analysis. Growth was performed in evacuated quartz cylindrical ampoules with conic tips (see Fig. S1 of Supporting Information file). Ampoule inner surface was graphitized. Each ampoule was placed into a furnace and equilibrated during a day at temperature 1-2°C higher than the nucleation temperature. Further the ampoule was pulled in the temperature gradient with the constant rate. Growth conditions are summarized in Tables 1 and S1 of Supporting Information file. Finally, our novel growth procedure allowed us obtain good quality bulk crystals of several cm in size and of whole range of composition in the systems under study.

### 3.2. Crystal growth of phases in quasi-binary systems

First, we discuss growth results for so called "124" compounds $GeBi_2Te_4$, $SnBi_2Te_4$, $PbBi_2Te_4$, $MnBi_2Te_4$, $MnSb_2Te_4$ in the corresponding quasi-binary systems. Respective to Fig. 1 A is GeTe, SnTe, PbTe or MnTe, ($M^1$=Ge, Sn, Pb, Mn), B is $Bi_2Te_3$, $Sb_2Te_4$ ($M^2$=Bi, Sb). These systems typically reveal complex phase relations including a number of compounds of general formula *n*A**m*B formed due to corresponding peritectic reactions (see Fig. 2, right column) [27–31]. Moreover, temperature ranges to grow individual phases are narrow, typically about 10 K. Therefore, it is rather difficult to grow crystal of certain phase from the melt.

To provide single crystal growth or, in other words, nucleation of minimal number of initial crystallites following conditions should be fulfilled: low overheating, minimal volume of overcooled melt, low pulling rate. First two conditions should be maintained during typical nucleation time of $10^3$-$10^4$ s [32]. The minimal volume is provided by ampoule geometry, which is a cylinder with a conic tip and a quartz heat sink rod attached (see Fig. S1 in SI file). Low melt overheating is required for nucleation, since in this case the melt structure which includes fragments with crystal-like local structure or chains of covalently bonded atoms. Melt overheating during nucleation is determined by the initial temperature gradient and the pulling rate. We have maintained the initial temperature gradient as 5-10 K/cm. The corresponding melt overheating, calculated as the difference between weighted mean temperature of the melt and temperature of the ampoule tip, was in the range of 2.5-6 K (Table 1). Under these conditions and the pulling rate lower than 0.4 mm/h the probability of single initial crystallite formation and further single crystal growth is about 50% as determined for $A^4B^6$ compounds [32].

For crystals obtained the crystallization front was typically perpendicular to layers direction (*c*-axes) similar to the case of $Bi_2Te_3$ and $Sb_2Te_3$, whereas for $Bi_2Se_3$ they are parallel to each other. This implies different nucleation and growth mechanisms. For $Bi_2Se_3$ layer-by-layer growth *via* 2D nucleation can be supposed, whereas for telluride compounds normal-like growth mechanism is reasonable to assume, where atoms are needed to migrate at the surface only within single septuple or quintuple layer (QL), i.e., not more than 1 nm.

Table 1 Growth conditions for three-phase-method. Nucleation temperature is measured by DTA, initial temperature gradient is calculated from the nucleation temperature and preliminary measured temperature profile in empty furnace, melt overheating is determined as the difference between mean melt temperature and ampoule tip temperature (nucleation)

| Sample | Nucleation temperature, ±0.2 °C | Initial tempera ture gra-dient, °C/cm | Melt overhea ting, °C | Pool-ing rate, mm/h | Stoichiometry of load | Composition of solid phase (tablet) | Load, g |
|---|---|---|---|---|---|---|---|
| $GeBi_2Te_4$_K | 584.9 | 6.2 | 3 | 0.25 | Ge 0.9995(4)<br>Bi 2.0004(0)<br>Te 4.0001(9) | GeTe | 15 |
| $SnBi_2Te_4$_K | 601.0 | 6.2 | 3.9 | 0.2 | Sn 1.0005<br>Bi 1.9996<br>Te 3.9999 | Sn 1.0001<br>Te 0.9999 | 10 |
| $MnBi_2Te_4$_K | 591.5 | 9.6 | 4 | 0.2 | Mn 1.0000(1)<br>Bi 2.0000(2)<br>Te 3.9999(3) | Mn 1.0000<br>Te 1.0000 | 15.6 |
| $MnSb_2Te_4$_K | 649.0 | 5.6 | 2.5 | 0.2 | Mn 1.0002(2)<br>Sb 1.9998(3)<br>Te 4.0000(5) | MnTe | 10 |
| $Ge_{0.4}Mn_{0.6}Bi_2Te_4$_K | 593.1 | 10.2 | 4 | 0.2 | Ge 0.404(3)<br>Mn 0.606(3)<br>Bi 1.993(4)<br>Te 4.000(6) | Ge 0.4001<br>Mn 0.6000<br>Te 0.9999 | 10 |
| $Ge_{0.2}Mn_{0.8}Bi_2Te_4$_K | 588.7 | 6.7 | 4 | 0.2 | Ge 0.1990<br>Mn 0.7974<br>Bi 2.0029<br>Te 4.0007 | Ge 0.1997<br>Mn 0.8003<br>Te 1.0000 | 10 |
| $Ge_{0.1}Mn_{0.9}Bi_2Te_4$_K | 592 | 9.8 | 3 | 0.2 | Ge 0.100(2)<br>Mn 0.90(3)<br>Bi 2.00(4)<br>Te 4.000(6) | Ge 0.100<br>Mn 0.900<br>Te 1.000 | - |
| $Ge_{0.05}Mn_{0.95}Bi_2Te_4$_K | 587 | 7.4 | 3.4 | 0.2 | Ge 0.049(2)<br>Mn 0.951(3)<br>Bi 2.00(4)<br>Te 4.000(6) | Ge 0.0500<br>Mn 0.9500<br>Te 1.0001 | 10 |
| $Ge_{0.025}Mn_{0.975}Bi_2Te_4$_K | 589 | 10.7 | 4 | 0.2 | Ge 0.023(3)<br>Mn 0.977(3)<br>Bi 2.00(4)<br>Te 4.000(4) | Ge 0.0252<br>Bi 0.9749<br>Te 0.9999 | - |
| $MnSb_{1.8}Bi_{0.2}Te_4$_K | 641.9 | 6.1 | 3 | 0.2 | Mn 1.04<br>Bi 0.19<br>Sb 1.77<br>Te 4.00 | MnTe | 15.1 |
| $MnSb_{1.2}Bi_{0.8}Te_4$_K | 620.2 | 7.1 | 6.1 | 0.2 | Mn 0.97<br>Bi 0.80<br>Sb 1.21<br>Te 3.97 | MnTe | 13.9 |
| $MnSb_{0.6}Bi_{1.4}Te_4$_K | 615.8 | 5.8 | 4.5 | 0.2 | Mn 0.99<br>Bi 1.41<br>Sb 0.60<br>Te 4.00 | MnTe | 13.9 |
| $MnSb_{0.5}Bi_{1.5}Te_4$_K | 603.4 | - | - | 0.2 | Mn 0.99<br>Bi 1.51<br>Sb 0.50<br>Te 4.00 | MnTe | 15 |
| $MnIn_{0.05}Bi_{1.95}Te_4$_K | 589.5 | - | - | 0.15 | Mn 0.9984<br>In 0.0510<br>Bi 1.9501<br>Te 4.0005 | Mn 0.9987<br>In 0.0510<br>Te 1.0753 | 14.7 |
| $MnIn_{0.1}Bi_{1.9}Te_4$_K | 587.5 | - | - | 0.15 | Mn 1.002<br>In 0.1002<br>Bi 1.8983<br>Te 3.9996 | Mn 1.0000<br>In 0.1000<br>Te 1.1500 | 15.5 |
| $MnIn_{0.15}Bi_{1.85}Te_4$_K | 591.5 | - | - | 0.15 | Mn 1.000<br>In 0.1499<br>Bi 1.8500<br>Te 3.9999 | Mn 1.0001<br>In 0.1499<br>Te 1.2250 | 14.9 |
| $MnIn_{0.5}Bi_{1.5}Te_4$_K | 591.7 | - | - | 0.15 | Mn 1.000<br>In 0.4998<br>Bi 1.5001<br>Te 4.0002 | Mn 1.0001<br>In 0.4999<br>Te 1.7500 | 14.9 |

To characterize crystal composition, in Fig. 2 we presented $M^1$/Te ratio as a universal indicator of crystal stoichiometry and structure. This is marked as horizontal bars. In Fig. 2a the data obtained for $GeBi_2Te_4$ grown by the three-phase method in comparison with Bridgman method. In the last case growth from 1:2:4 stoichiometric melt results in primary crystallization of another compound, $Ge_2Bi_2Te_5$, yielding from peritectic reaction $p_1$ in the equilibrium phase diagram shown in the left panel of Fig. 2a. In the case of three-phase growth we used GeTe tablet and $Bi_2Te_3$ melt, with the figurative point corresponding to 1:2:4 stoichiometry (50 mol.% GeTe). Although it formally gives the same equilibrium conditions as they are in Bridgman growth, 225-compound forms at the tablet surface, that creates a local equilibrium between 225-phase and the melt enriched in $Bi_2Te_3$. Such melt can crystallize with direct formation of 124-phase in ampoule tip due to peritectic reaction $p_2$. Therefore, whole ingot includes solely 124-phase in this case. In the case of $SnBi_2Te_4$ Bridgman growth also results in multiphase ingot, with the initial part containing targeted 124 phase and final part being 147-phase. This behavior can be related to some kinetic reasons such as low natural crystallization rate for this system. For three-phase method performed with slightly lower pulling rate whole ingot includes solely 124-phase. In the case of $PbBi_2Te_4$ the above-mentioned condition to apply the three-phase method in vertical arrangement is violated from the higher density of PbTe than that of melt. For this reason, PbTe tablet should drown that disturbs crystal nucleation. At the same time, Bridgman growth seems to be successful in this case.

For Mn-containing systems which are the most interesting for applications as magnetic topological insulators Bridgman growth totally fails in accordance with previous reports [33], whereas the three-phase growth produces bulk crystals of several cm in size (in our case, generally crystals can be much larger). Crystals obtained typically present a sequence of 124-, 147- and 1610-phases grown one by one, that is clearly seen in Fig. 2d,e. In comparison with $GeBi_2Te_4$ and $SnBi_2Te_4$ cases, the equilibrium phase diagrams for Mn-containing systems have steeper liquidus that creates generally lower supersaturation when crystallization temperature decreases due to the delay of the natural growth rate and fast pulling. Therefore, this factor acts in opposite manner and should be favorable with respect to crystal quality. Another possible reason is slow tablet dissolution. Although dissolution rate generally should be higher than the crystallization one, in the case of MnTe it can be lowered due to higher reactivity of MnTe and hence thicker oxide layer covering the tablet. If the dissolution is slow the melt is depleted in Mn and becomes nutrient phase for 147- and 1610-crystals. Therefore, one can conclude that the tablet dissolution rate is one of the key factors influencing crystal quality.

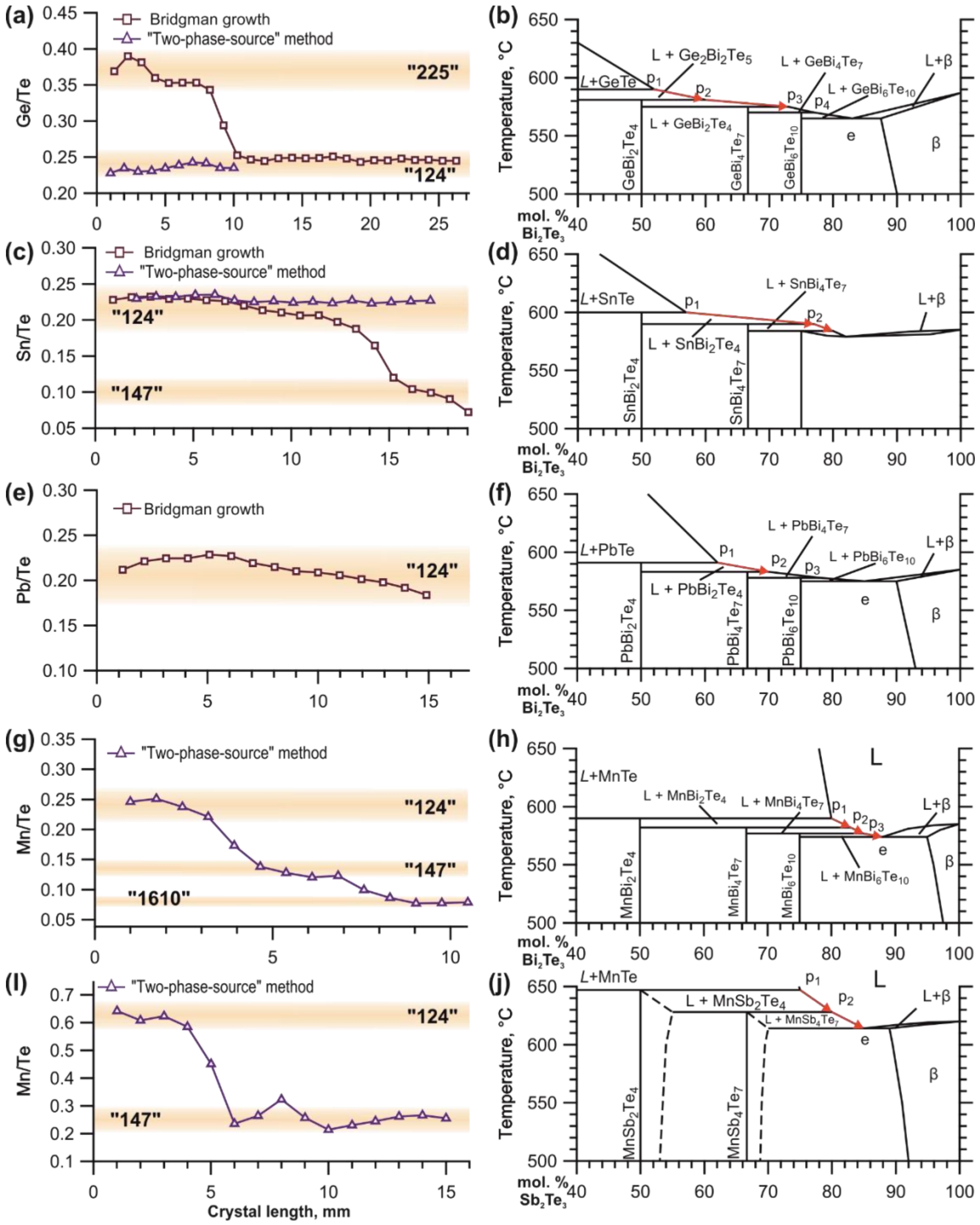

Fig.2. Crystal growth of phases in quasi-binary systems $GeBi_2Te_4$, $SnBi_2Te_4$, $PbBi_2Te_4$, $MnBi_2Te_4$, $MnSb_2Te_4$: $M^1$/Te atomic ratio ($M^1$ =Ge,Sn,Pb,Mn) in crystal obtained (left) and corresponding equilibrium phase diagrams (right) [27–31]

Crystal quality is illustrated in Fig. 3 and Figs. S2-S5 (see Supporting Information file) with the example of $MnBi_2Te_4$. Fig.3a shows HAADF-STEM image and corresponding EDX maps for Mn, Bi and Te evidencing structure composed by septuple layers (SLs) and preferential but not complete occupancy of the central layer of SL by Mn. The presence of antisite $Bi_{Mn}$ and $Mn_{Bi}$ defects is corroborated by QSTEM simulation results shown in Fig. S4. Cleavage of the crystal along (111) plane creates a flat surface with low density of steps of typical height corresponding to 1-2 SL. AFM image obtained near steps is exhibited in Fig. 3b with the corresponding line profile. Atomic-resolution STM image is shown in Fig. 3c. Although the surface is atomically flat electronic effects due to defects (for instance, antisites) create an additional contrast. Fourier transformation (FFT) of surface topography in inset includes mainly quasiparticle interference (QPI) features that originates from elastic scattering of surface electrons on the point defects and in general confirms high structural perfection in the presence of topological surface states (TSS).

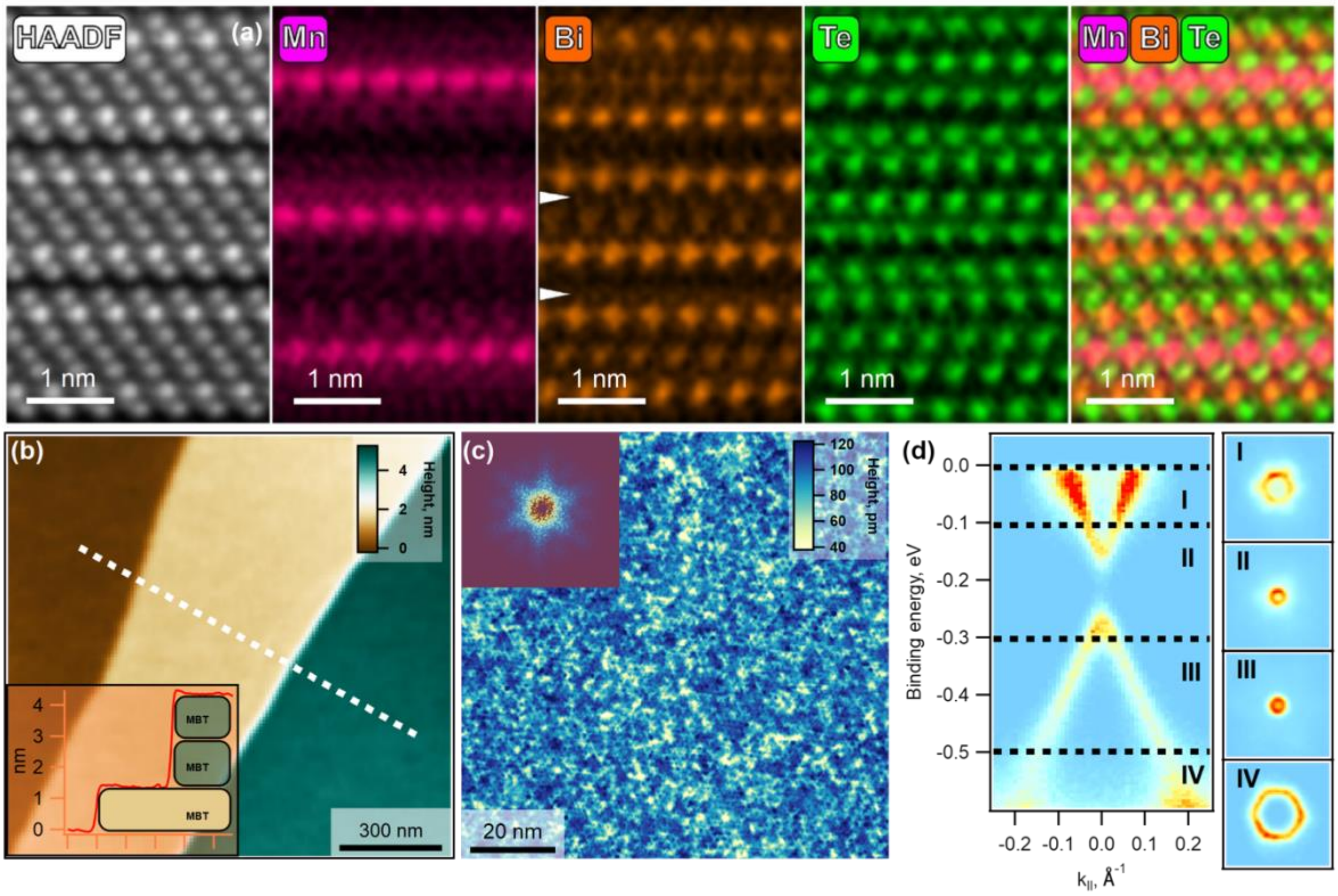

Fig.3. Characterization of $MnBi_2Te_4$ crystal: (a) cross-sectional HAADF STEM image and atomic-resolution EDX maps of the same area, (b) AFM image, (c) STM image taken at U = 100 mV, I=900 pA. Corresponding FFT image is shown as inset. (d) ARPES spectra taken along *Γ-K* direction at hν=21.2 eV

Band structure along the *K-Γ-K* direction of the surface Brillouin zone measured by ARPES at photon energy of 21.2 eV in Fig. 3d demonstrates sharp features evidencing high crystal perfection. At this photon energy the photoemission (PE) cross-section for the (TSS) of $MnBi_2Te_4$ is rather low, and PE signal from the bulk bands dominates in the spectra [5]. The bulk valence and conduction bands have a nearly cone-like shape. Its constant-energy contours are represented by hexagonally warped circles with a decrease of warping close to the top/bottom of the bands. According to Hall measurements bulk carrier concentration $n$=0.93·$10^{20}$ $cm^{-3}$ and mobility $\mu$=65 $cm^2$/(V·s) at 4.7K. Neel temperature is 24K.

### 3.3. Growth of crystal with cation substitution in quasi-ternary systems

For researches and device applications, it is needed to prepare mixed crystals of solid solutions to tune finely their electronic band structure and properties. Especially this concerns Mn-containing crystals. For solid solutions band structure is a function of mole fraction, which should be, therefore, controlled precisely during crystal growth. However, compare to the individual 124-compounds, phase equilibria near quasi-binary sections which tie two 124-phases in the corresponding quasi-ternary systems are much more complicated and almost unstudied. Besides, larger number of components creates more freedom for the system from thermodynamic point of view. Restricting the freedom by increasing number of phases, like in our three-phase method, is also beneficial for better control of crystal composition.

We have performed growth of $Mn(Bi,Sb)_2Te_4$, $(Ge,Mn)Bi_2Te_4$ and $Mn(Bi,In)_2Te_4$ mixed crystals using three-phase method. For this we used the source initially included two phases $L((Bi,Sb)_2Te_3)+S(MnTe)$, $L(Bi_2Te_3)+S(Ge_xMn_{1-x}Te)$ or $L(Bi_2Te_3)+S(In_xMn_{1-x}Te)$ correspondingly. Growth conditions are summarized in Table 1. The two-phase source method resulted in good-quality, several-cm-sized crystals of whole composition range of these systems. It should be noted that classical Bridgman growth was successful only for $(Ge,Mn)Bi_2Te_4$ mixed crystals of low Mn concentration (Table S1). For $Mn(Bi,Sb)_2Te_4$ and $Mn(Bi,In)_2Te_4$ this procedure does not produce single crystals.

Some of the typical crystals obtained and their compositions are exhibited in Fig. 4. Most of the experiments show the general tendency of (Ge+Mn)/Te ratio for mixed crystals in Fig. 4 to decrease during crystal growth similarly as it was observed for pure $MnBi_2Te_4$ in Fig. 2. This composition behavior corresponds to appearance of different crystal structures composed of SLs and QLs. The structures are marked with horizontal bars in Fig. 4 left. The underlying reason of such behavior seems to be the same, namely low dissolution rate for tablets. This supposition is supported by the fact that the natural growth rate should be lower for mixed crystals, but switching from 124-phase to 147-phase takes place at nearly same length for quasi-binary and quasi-ternary systems.

Middle column in Fig. 4 presents mole fraction $x$ distribution along the crystals of solid solutions. In the case of $Mn(Bi_{1-x}Sb_x)_2Te_4$ crystals of all compositions are nearly uniform in $x$ (Fig. 4b). Correlation between the first crystal composition (1 mm part in ampoule tip) *vs* source composition is plotted in Fig. 4c with red cycles. The effective segregation coefficient $\alpha$, defined as a ratio of the first crystal composition $x$ to integral initial composition of the source, seems to be slightly higher than unity. Therefore, crystals are slightly enriched in $MnSb_2Te_4$ component. A possible reason is related to the fact that the crystallization path lies along the maximal slope of solidus surface in quasi-ternary system $MnTe-Bi_2Te_3-Sb_2Te_3$. At the same time the direction of the maximal slope may deviate from the line linking the corresponding figurative point and MnTe corner.

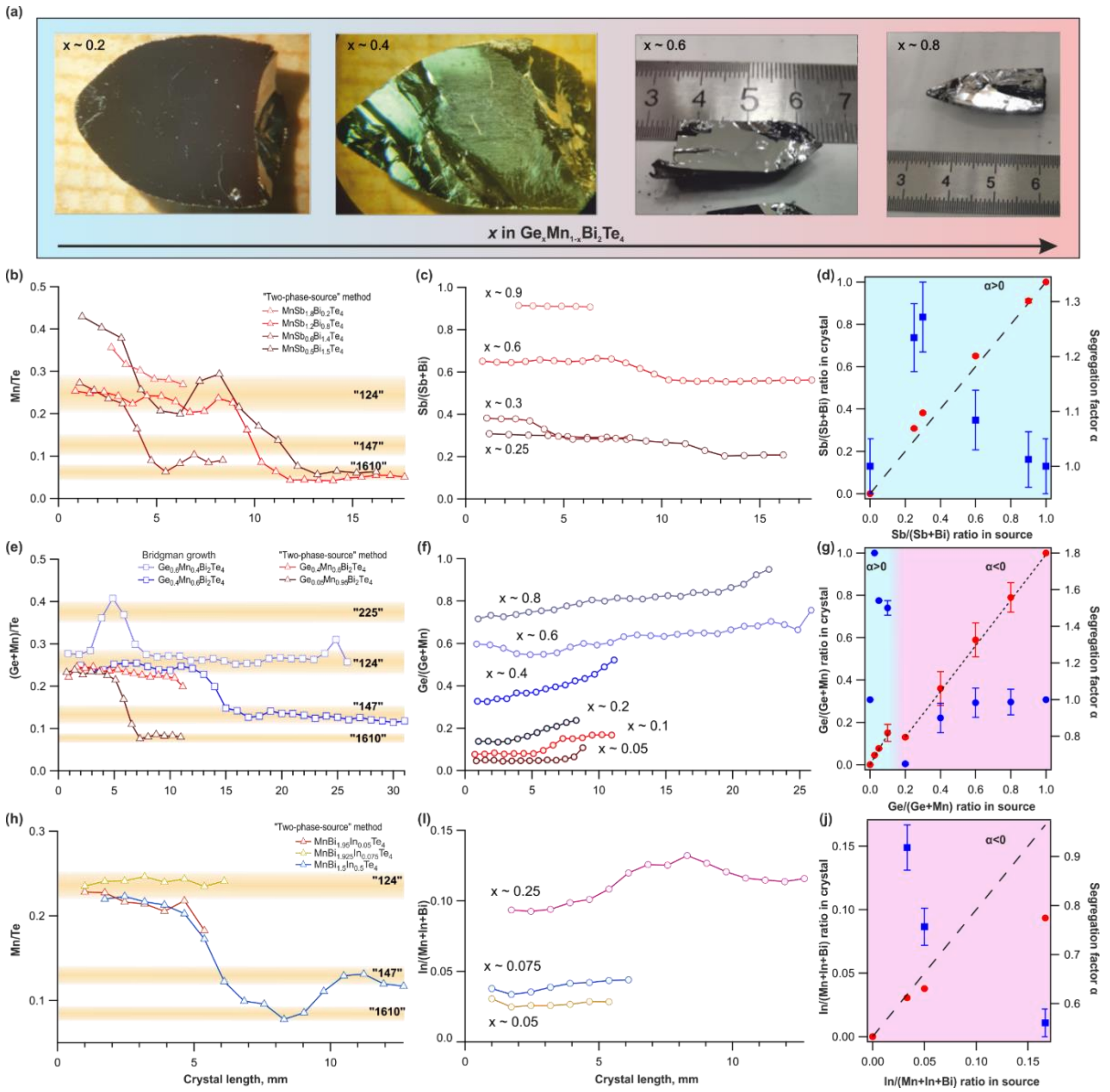

Fig. 4. Crystal growth of solid solutions Mn(Bi,Sb)$_2$Te$_4$, (Ge,Mn)Bi$_2$Te$_4$ and Mn(Bi,In)$_2$Te$_4$: general view of some (Ge,Mn)Bi$_2$Te$_4$ crystals (upper line), M$^1$/Te atomic ratio (M$^1$=Ge,Sn,Pb,Mn) in crystal obtained (left), mole fraction distribution along the crystal (middle), and effective segregation coefficient (right)

For Ge$_x$Mn$_{1-x}$Bi$_2$Te$_4$ with $x$≥0.6 massive bulk crystals of satisfactory quality can be obtained by common Bridgman method. At lower $x$, two-phase source enables high quality bulk crystals. In all cases, mole fraction $x$ increases upon growth, whereas the plot of effective segregation coefficient splits into two parts at $x$=0.2. At x<0.2 α>1, at x>0.2 α is slightly lower that unity. It is probably related to the phase relations in the tablet of Ge$_x$Mn$_{1-x}$Te. In detail, there is demixing in GeTe-MnTe system for $x$=0.06-0.038 (see Fig. S6).

In the case of Mn(Bi,In)$_2$Te$_4$ In substitute both Bi and Mn. For this reason, atomic ratio In/(Mn+In+Bi) is plotted in Fig. 4h. The effective segregation coefficient seems to be below unity, indicating less indium in crystals as compared to the source. Indium content slightly increases along crystals.

Crystal quality is illustrated in Fig. 5 with the example of $Mn(Bi_{1-x}Sb_x)_2Te_4$. Figs. 5a show cross-sectional HAADF-STEM image for the sample with $x$=0.9, its fragment and corresponding EDX maps for Mn, Bi and Te evidencing structure composed by SLs, random mixing of Bi and Sb, and preferential but not complete occupancy of the central layer of SL by Mn. XRD data are provided in Figs. S6, S7 of the Supporting Information File. The variation of band structure along for different compositions is illustrated in Fig. 5b. The spectra are plotted along the *K*-*Γ*-*K* direction of the surface Brillouin zone measured by ARPES at photon energy of 21.2 eV. They demonstrate quite sharp features evidencing high crystal perfection. At this photon energy the photoemission (PE) cross section for the topological surface states (TSS) of $MnBi_2Te_4$ is rather low, and PE signal from the bulk bands dominates in the spectra. The bulk valence ($BVB^1$) and conduction (BCB) bands have a nearly cone-like shape (V-shape). Also, M-shape is clearly seen as $BVB^2$. Crystal with $x$=0.38 has FL inside band gap. Fermi level position relative to bulk bands and Hall measurements in Fig. 5c demonstrates smooth variation. Bulk carrier concentration varies from $n=0.93\cdot10^{20}$ $cm^{-3}$ to $p=1.8\cdot10^{20}$ $cm^{-3}$ *via* minimum corresponding to $4.7\cdot10^{18}$ $cm^{-3}$. Cleavage of the crystal along (111) plane creates flat surface with low density of steps with typical height of 1-2 SLs. MFM images for crystal with $x$=0.9 are presented in Fig. 5d for two temperatures, 34K and 4K, corresponding to paramagnetic and ferromagnetic phases. Magnetic domains are clearly seen below Curie temperature (25K). Magnetic measurements demonstrate magnetic hysteresis. SQUID magnetometry data are provided in Fig. S8 of the Supporting Information File.

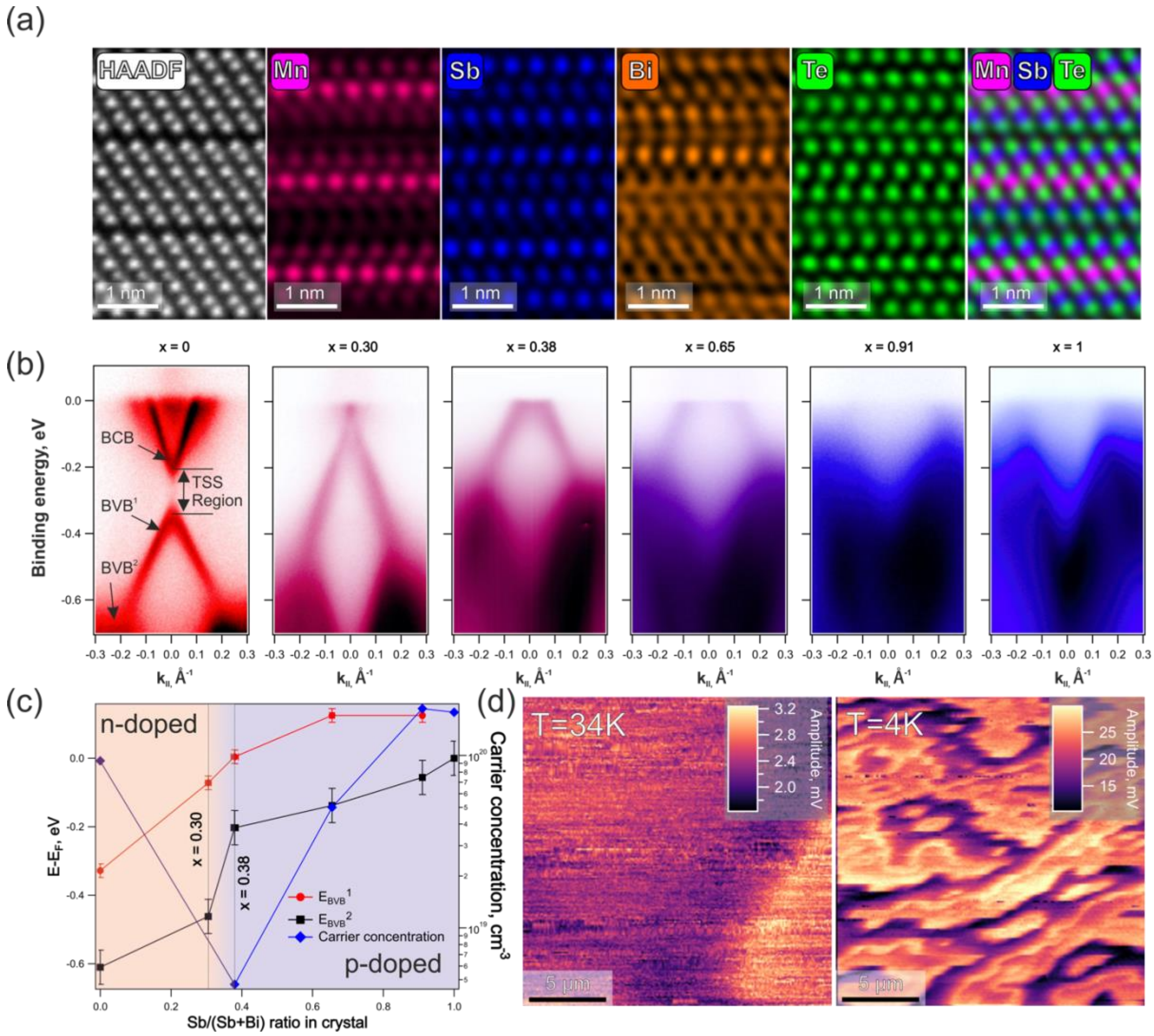

Fig. 5. Characterization of magnetic topological insulator $Mn(Bi,Sb)_2Te_4$: (a) cross-sectional HAADF STEM image and elemental EDX maps of the same area, (b) ARPES dispersion obtained in *Γ-K* direction measured at *hv*=21.2 eV for samples of different composition as indicated, (c) the composition dependence of Fermi level position relative to V-shape-subband ($BVB^1$) and M-shape-subband ($BVB^2$) of valent band and the corresponding carrier concentration measured at 4K (with exception of *x*=0.9 measured at 60 K), (d) MFM images above and below Curie temperature, showing ferromagnetic domains.

## Conclusions

In summary, we have developed new procedure to grow large-scale bulk crystals of incongruently melted compounds and solid solutions which enables preparation of several magnetic topological insulators. In our method coexistence of following three-phases in the growth system is maintained: growing crystal, melt and solid high-melting component as flowing tablet. This restricts system freedom and provides high quality crystals of $GeBi_2Te_4$, $SnBi_2Te_4$, $PbBi_2Te_4$ homogeneous in composition. For Mn- containing compounds ($MnBi_2Te_4$, $MnSb_2Te_4$) and solid solutions ($Mn(Bi,Sb)_2Te_4$, $(Ge,Mn)Bi_2Te_4$ and $Mn(Bi,In)_2Te_4$) we observed sequential formation of 124, 147 and 1610 phases, most probably due to low dissolution rate of tablets. To provide growth of solely 124-phase in the case of Mn-containing systems special precautions should be undertaken to avoid pollution (oxidation) of initial Mn-containing material, that influences its dissolution rate in melt.

### Acknowledgments

We are grateful to Russian Science Foundation for financial support (Grant 23-72-00020). N.V. Vladimirova acknowledges RFBR individual support within PhD project (grant 20-33-90273). A.S.F. thank the Ministry of Science and Higher Education of the Russian Federation (Grant No. 075-15-2024-632) for the grant in the form of a subsidy for the purpose of implementing a scientific project under the guidance of an invited leading scientist.

## Figure captions

Fig.1. Schematic representation of incongruently melted compound growth using two-phase source: a fragment of *T-x* phase diagram (left) and corresponding positions of growth ampoule relative to a temperature profile during different steps (nucleation, growth, termination).

Fig.2. Crystal growth of phases in quasi-binary systems $GeBi_2Te_4$, $SnBi_2Te_4$, $PbBi_2Te_4$, $MnBi_2Te_4$, $MnSb_2Te_4$: $M^1$/Te atomic ratio ($M^1$ =Ge,Sn,Pb,Mn) in crystal obtained (left) and corresponding equilibrium phase diagrams (right) [27–31]

Fig.3. Characterization of $MnBi_2Te_4$ crystal: (a) cross-sectional HAADF STEM image and atomic-resolution EDX maps of the same area, (b) AFM image, (c) STM image taken at U = 100 mV, I=900 pA. Corresponding FFT image is shown as inset. (d) ARPES spectra taken along *Γ-K* direction at hv=21.2 eV

Fig. 4. Crystal growth of solid solutions $Mn(Bi,Sb)_2Te_4$, $(Ge,Mn)Bi_2Te_4$ and $Mn(Bi,In)_2Te_4$: general view of some $(Ge,Mn)Bi_2Te_4$ crystals (upper line), $M^1$/Te atomic ratio ($M^1$ =Ge,Sn,Pb,Mn) in crystal obtained (left), mole fraction distribution along the crystal (middle), and effective segregation coefficient (right)

Fig. 5. Characterization of magnetic topological insulator $Mn(Bi,Sb)_2Te_4$: (a) cross-sectional HAADF STEM image and elemental EDX maps of the same area, (b) ARPES dispersion obtained in *Γ-K* direction measured at *hv*=21.2 eV for samples of different composition as indicated, (c) the composition dependence of Fermi level position relative to V-shape-subband ($BVB^1$) and M-shape-subband ($BVB^2$) of valent band and the corresponding carrier concentration measured at 4K (with exception of $x$=0.9 measured at 60 K), (d) MFM images above and below Currier temperature, showing ferromagnetic domains.